\documentclass[twocolumn,showpacs,preprintnumbers,amsmath,amssymb,prl,aps]{revtex4-1}
\usepackage{epsfig}% Include figure files
\usepackage{graphicx}% Include figure files
\usepackage{dcolumn}% Align table columns on decimal point
\usepackage{bm}% bold math
\usepackage{float}
\usepackage{multirow}

\begin{document}

\title{Anomalies in the electronic structure of Bi$_2$Se$_3$}

\author{Deepnarayan Biswas}
\author{Kalobaran Maiti\footnote{Corresponding author: kbmaiti@tifr.res.in}}
\affiliation{Department of Condensed Matter Physics and Materials'
Science, Tata Institute of Fundamental Research, Homi Bhabha Road,
Colaba, Mumbai - 400 005, India.}

\date{\today}

\begin{abstract}
We studied the electronic structure Bi$_2$Se$_3$ employing density
functional theory. The calculations show that the Dirac states
primarily consists of the states at the interface of surface and
sub-surface quintuple layers and the emergence of the Dirac states
depends on the surface terminations in sharp contrast to their
surface character expected for such systems. This manifests the
complexity of real materials and the realization of topological
order in real materials as an outstanding problem. We discover that
the surface character of the Dirac states can be achieved by
adsorption of oxygen on Bi-terminated surface due to the change in
covalency by the relatively more electronegative oxygens.
\end{abstract}


\maketitle

%\section{Introduction}

Topological insulators are characterized by metallic states at the
surface of bulk insulators protected by time reversal symmetry and
strong spin-orbit coupling \cite{hasan_rev,3dti}. Due to time
reversal symmetry protection, the electrons moving on the surface
cannot be scattered in the reverse direction without spin flip.
Thus, the corresponding energy bands exhibit linear dispersion
forming a Dirac cone, which is one of the key signatures in the
electronic structure of a topological insulator. Due to such unique
electronic properties, these materials are potential candidates for
realization of exotic physics such as Majorana Fermions, magnetic
monopoles \cite{monopole,majorana}. These materials are also
expected to bring immense technological advances and new
possibilities in the fields of spintronics, quantum computation,
dissipation less charge transfer etc. \cite{3dti}. While all these
realizations require experimental finding of protected topologically
ordered surface states on insulating bulk materials, experiments
show contrasting scenarios often with instability of topological
order. The bulk of almost all the materials studied exhibit metallic
phase \cite{aging_hasan}. Thus, finding of a true topological
insulator remains to be an outstanding puzzle.

\begin{figure}%[H]
 %\vspace{5ex}
 \includegraphics[scale=.4, trim=-0cm 0cm 0cm 0cm,clip]{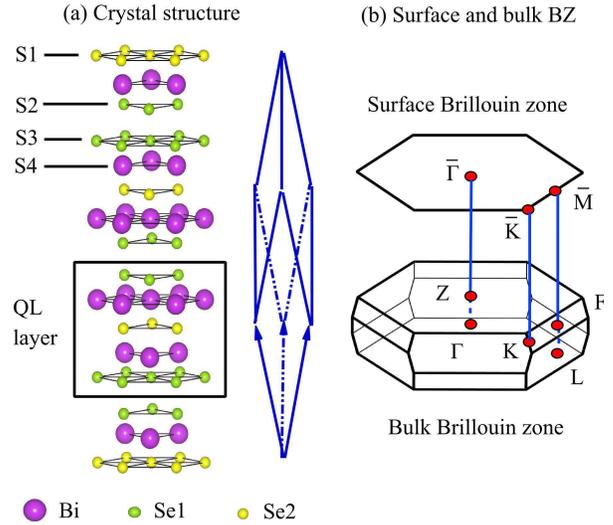}  %trim option's parameter order: left bottom right top
 \vspace{-2ex}
\caption{(a) Crystal structure of Bi$_2$Se$_3$ exhibiting atomic
arrangement and the rhombohedral unit cell. S1, S2, S3 and S4 denote
the layers where the surface terminations are considered in this
work. (b) The surface and bulk Brillouin zone (BZ) along with
different high symmetry points.}
 \vspace{-2ex}
\end{figure}

Engineering new materials for such behavior requires critical
understanding of the electronic structure of the materials
exhibiting topological order. For example, one of the most studied
materials for such behavior is Bi$_2$Se$_3$ that shows Dirac cone
with its apex at finite binding energy instead of its chemical
potential. In addition, several bulk bands are observed at the Fermi
level suggesting its metallic bulk electronic structure. Carrier
doping due to aging/impurities often inhibits achieving topological
transport regime. Some theories predicted relaxation of van der
Waals bond as a reason for such behavior \cite{relaxation_noh},
while other studies show necessity of unusually large change in bond
length to produce the experimentally observed scenario
\cite{dft_fukai}. The surface band bending induced by adsorption of
residual gases \cite{BB_bianchi,dft_wang,BB_zhang,deep}, commonly
observed in semiconductors, may also give rise to such effect.
Interestingly, both positive and negative charge doping due to
impurity adsorption have been observed in angle resolved
photoemission measurements \cite{deep,o2_shen,aging_kong}. While all
these studies indicate instability of the surface states with aging,
impurity adsorption etc., there are studies reporting stable surface
behavior \cite{contradict_rader,stable_atuchin,stable_goly}.
Evidently, the behavior of the surface states of these materials is
highly anomalous. In order to reveal microscopic details underlying
such electronic properties, we calculated the electronic band
structure for both surface and bulk employing density functional
theory, and discover interesting results, significantly different
from the expected behavior of this system.

The electronic band structure calculation of Bi$_2$Se$_3$ was
carried out using full potential linearized augmented plane wave
method within the local density approximations \cite{wien2k}. The
bulk electronic structure is calculated using the lattice constants,
$a$ = $b$ = 4.18~\AA, $c$ = 28.7~\AA, $\alpha$ = $\beta$ = 90$^o$
and $\gamma$ = 120$^o$ obtained from the literature \cite{lattice}.
In order to investigate the surface electronic structure, we
considered the lattice in a 'slab' configuration with at least 5
quintuple layers within the unit cell - the total number of layers
varies for different surface terminations. Calculations were carried
out for all the possible surface terminations as defined in Fig. 1.

Bi$_2$Se$_3$ forms in rhombohedral structure with space group,
$R\bar{3}m$. In Fig. 1(a), we show the crystal structure\cite{vesta}
with both hexagonal and rhombohedral axes. It has a layered
structure containing quintuple layers consisting of
Se1-Bi-Se2-Bi-Se1 layers, where two nonequivalent Se layers are
denoted by Se1 and Se2. The quintuple layers are believed to be
connected through weak Van der Waals bond \cite{relaxation_noh,
dft_fukai}; the inter-layer bonding within the quintuple layer is
relatively stronger. This indicates a well defined cleavage plane
exposing Se1 terminated surface. Recent experiments, however, show
signature of both Bi and Se terminated surfaces on cleaving
\cite{deep,tdas_stm} exhibiting significantly different behavior.
Thus, we considered the surface terminations S1, S2, S3 and S4 (see
figure) representing Se2, Se1 (one Se1 on top of the quintuple
layer), Se1 (normal cleavage plane) and Bi terminations,
respectively. Fig. 1(b) shows the bulk and surface Brillouin zones
with different high symmetry $k$ points. For (111) surface of
Bi$_2$Se$_3$, the time reversal invariant momenta (TRIM) are the
high symmetry points, $\Gamma$ and $M$. In addition to the pristine
cases, we deposited a monolayer of oxygen on top of the surfaces.
The exchange correlation potential is included under generalized
gradient approximation (GGA) \cite{GGA}. Spin-orbit interaction is
included as a second order perturbation for Bi and Se atoms
\cite{yeir2o7_band}. The energy convergence was achieved using
$10\times 10\times 1$ $k$-mesh.

\begin{widetext}

\begin{figure}%[H]
 %\vspace{5ex}
 \includegraphics[scale=.7, trim=0cm 0cm -1cm 0cm,clip]{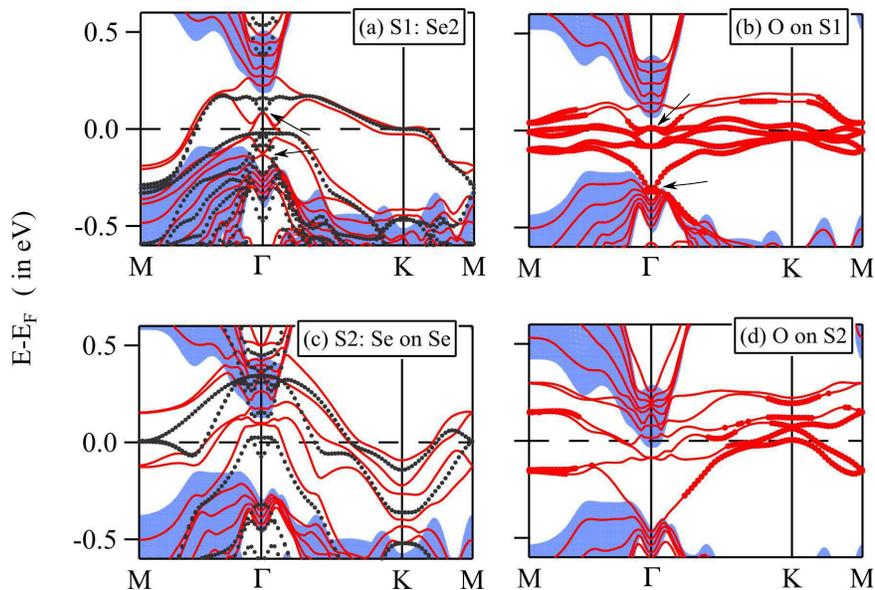}  %trim option's parameter order: left bottom right top
 \vspace{-2ex}
\caption{Band structure of (a) clean and (b) 1 monolayer (ML) O
covered S1 terminated surfaces, and (c) clean and (d) 1 ML O covered
S2 terminated surface. The black dots show the bands with no
spin-orbit coupling and lines are with spin-orbit coupling. The blue
region is the projected bulk bands and the red dots denote the bands
with more than 50\% O contribution.}
 \vspace{-2ex}
\end{figure}

\end{widetext}

%\section{Results}

The calculations for the bulk Bi$_2$Se$_3$ converge to insulating
ground state with a band gap of about 0.3 eV consistent with the
experimental findings
\cite{bandgap_black,bandgap_mooser,band_inversion}. The results from
the slab calculations exhibit interesting scenario with very
different surface bands near the Fermi level, $\epsilon_F$ for
different terminations. All the energy bands obtained from slab
calculations are doubly degenerate as a symmetric slab geometry has
been used \cite{symm_slab} for these calculations. In Figs. 2(a) and
2(c), we show the band structure obtained for S1 and S2 cases. The
black dots represent the bands without spin-orbit coupling (SOC) and
the solid red lines correspond to the calculations including SOC.
The shaded area represents the projected bulk bands (PBB) for the
SOC case, where the difference in Fermi level between the surface
and bulk calculations are compensated by matching the bulk bands
away from the Fermi level. The bands appearing in the gap of the PBB
are the surface bands. It is evident from the data that SOC leads to
a splitting of the spin degenerate bands obtained without SOC
\cite{yeir2o7_band}. For both the surface terminations, the surface
bands cross the Fermi level between surface projected TRIMs,
$\Gamma$-M odd number of times implying that the material belong to
a strong topological insulator class \cite{3dti}. The bands near
$\epsilon_F$ consist primarily of Se 4$p$ and Bi 6$p$ characters.

For clean Se2 terminated surface, the conduction band maxima and the
valence band minima at $\Gamma$ point shown by arrows in Fig. 2(a)
exhibit a separation of about 0.3 eV due to Bi $p$ and Se2 $p$
hybridizations. The band crossing at around -0.16 eV exhibits a
scenario akin to a Dirac cone representing topological order. Here,
the finite energy of the Dirac point (apex of the cone) is intrinsic
to the electronic structure. Impurities/defects play important role
in the electronic structure leading to charge carrier doping and
introducing local character of the charge carrier
\cite{cab6,cab6band}. We have verified this in the present case by
calculating electronic structure with one monolayer oxygen on the
surface. The Dirac point shifts away from $\epsilon_F$ as shown by
an arrow in Fig. 2(b). An additional pair of bands appear near
$\epsilon_F$ possessing primarily O 2$p$ character. This is shown by
big circles representing more than 50\% O 2$p$ contributions in Fig.
2(b). The energy bands near $\epsilon_F$ becomes narrower compared
to the pristine case suggesting enhancement of local character of
the electronic states. The overall shift of the energy bands towards
higher binding energies (= negative of the energy shown in the
figures) in the later case represents an effective electron doping
as expected due to relatively higher electronegativity of oxygen.
The results for S2 termination are shown in Fig. 2(c) and 2(d). The
energy bands exhibit significantly different behavior with no
signature of Dirac cone. Oxygen in this case leads to a band
narrowing (see Fig. 2(d)) and effective electron doping as found in
the case of S1 termination. However, the oxygen 2$p$ character
appears stronger around high symmetry points other than $\Gamma$.

\begin{widetext}

\begin{figure}%[H]
% \vspace{5ex}
 \includegraphics[scale=.7, trim=0cm 0cm 0cm 2.5cm,clip]{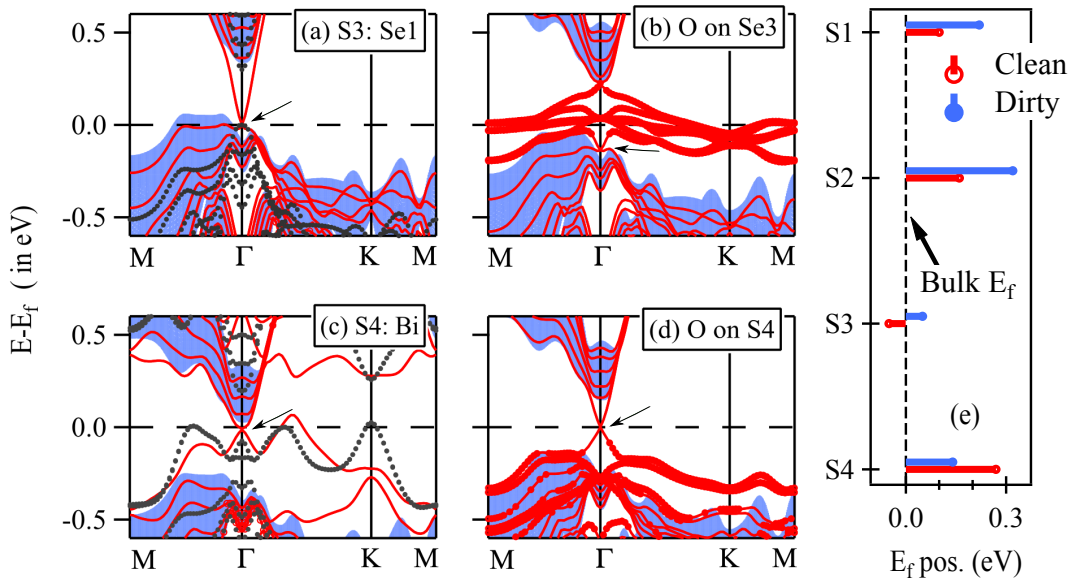}  %trim option's parameter order: left bottom right top
 \vspace{-2ex}
\caption{Energy band structure for (a) S3-terminated, (b) 1 ML
oxygen on S3 surface, (c) S4 terminated surface and (d) 1 ML oxygen
on S4 surface. (e) Shift of slab Fermi level w.r.t. bulk Fermi level
(dotted line) binding energy.}
 \vspace{-2ex}
\end{figure}

\end{widetext}

The energy bands corresponding to S3 and S4 terminated surfaces are
shown in Fig. 3. The results obtained without SOC exhibit finite gap
between the valence and conduction bands. Inclusion of SOC leads to
the formation of distinct Dirac cone like feature at $\Gamma$ point
at $\epsilon_F$ as shown by arrow in the figure. In these
terminations, the presence of spin-orbit coupling leads to the
surface bands crossing the Fermi level odd number of times between
TRIMs, $\Gamma$ and $M$ providing again the signature of a strong
topological insulating phase in these cases
\cite{tdas_stm,band_inversion}. The Dirac point (DP) appears very
close to the top of the valence band for Se terminated surface (S3
case) and near bottom of the conduction band on Bi-terminated
surface (S4 case).

The most interesting phenomena occurs on deposition of oxygen on
these surfaces; the Dirac cone shifts to higher binding energy
($\sim$~0.14 eV) along with significant narrowing of the energy
bands near $\epsilon_F$ (see Fig. 3(b)). O 2$p$-derived additional
narrow bands appear in the vicinity of the Fermi level as found in
the case of Se2 terminated surface. Fig. 3(d) exhibits the energy
bands corresponding to a monolayer oxygen on Bi-terminated surface.
The DP shifts away from the bottom of the conduction band and the
Dirac cone becomes prominent, better defined with larger slope.
Here, the oxygen 2$p$-derived bands with smaller dispersion appear
below $\epsilon_F$ unlike the Se-terminated cases. This makes the
topological surface states better accessible and easily discernible
from the impurity bands.

In Fig. 3(e), we show the Fermi energy found in the slab
calculations relative to that in the bulk calculation. While S1, S2
and S4 cases show an increase in Fermi energy, the Fermi energy
becomes smaller than the bulk Fermi energy in S3 case. The S4 case
(Bi terminated surface) exhibits the highest energy difference.
Presence of 1 ML oxygen leads to a shift of the Fermi level towards
higher energies suggesting an effective electron doping into the
system for all the Se terminations. Ironically, the deposition of
oxygen on Bi terminated surface (S4 case) leads to a shift of the
Fermi level in opposite direction - towards lower energies
suggesting an effective hole doping in this case consistent with the
experimental observations \cite{deep}.

%\section{Discussion}

%\begin{widetext}

\begin{figure}%[H]
% \vspace{5ex}
 \includegraphics[scale=.45, trim=-.8cm 0cm 0cm 0cm,clip]{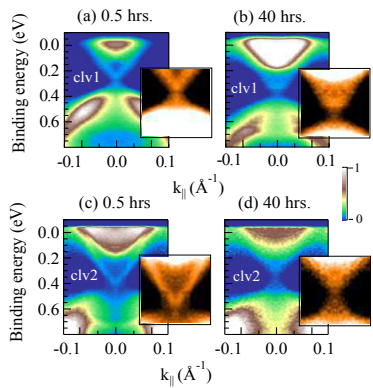}  %trim option's parameter order: left bottom right top
 \vspace{-2ex}
\caption{Angle resolved photoemission (ARPES) data adapted from
ref.\cite{deep}. ARPES data from Se-terminated surface (a) freshly
cleaved and (b) 40 hours aged, and Bi-terminated surface (c) freshly
cleaved and (d) 40 hours aged. Inset shows Dirac cone region with
magnified intensity.}
 \vspace{-2ex}
\end{figure}

%\end{widetext}

It is clear that the electronic structure corresponding to different
surface terminations of the system exhibit different scenario. The
sample with Se1 bi-layer on top (S2 case) do not show Dirac cone
although there are odd number of fermi level crossing of the energy
bands between TRIMs. Here the top Se layers are weakly coupled with
the bulk quintuple layer underneath and the surface electronic
structure possess effectively two dimensional nature.

The samples with Se2, Se1 and Bi terminated surface as discussed in
S1, S3 and S4 cases exhibit Dirac cone in their energy band
structure. The presence of a layer of oxygen leads to instability of
the Dirac states corresponding to the Se-terminated surface. In all
the cases, the surface oxygen bands appear close to the Fermi level
along with a shift of the Dirac cone to higher binding energies
\cite{deep,aging_kong}. Interestingly, the Bi-terminated surface
exhibit more prominent Dirac states on oxygen deposited surface. In
order to investigate the correspondence of this finding with the
experimental scenario, we show the experimental angle resolved
photoemission results in Fig. 4 adapted from the Ref.\cite{deep}.
The experimental results for the Se-terminated (Clv1) and
Bi-terminated (Clv2) cases are shown in upper and lower panels of
Fig. 4, respectively. The Freshly cleaved surface exhibit signature
of Dirac cone in both the cases as expected. Oxygen accumulation on
the Se-terminated surface leads to a shift of the Dirac point
towards higher binding energy due to effective electron doping as
found in this study too, along with a weakly defined Dirac cone.
However, the Bi-terminated surface exhibits energy shift in opposite
direction as predicted in Fig. 3(e) and exhibits a better defined
conical shape of the Dirac cone. These results clearly corroborate
the theoretical description shown in Fig. 3 establishing the robust
nature of the Dirac states on Bi-terminated surface.

\begin{figure}%[H]
% \vspace{5ex}
 \includegraphics[scale=.35, trim=0cm 0cm 0cm 0cm,clip]{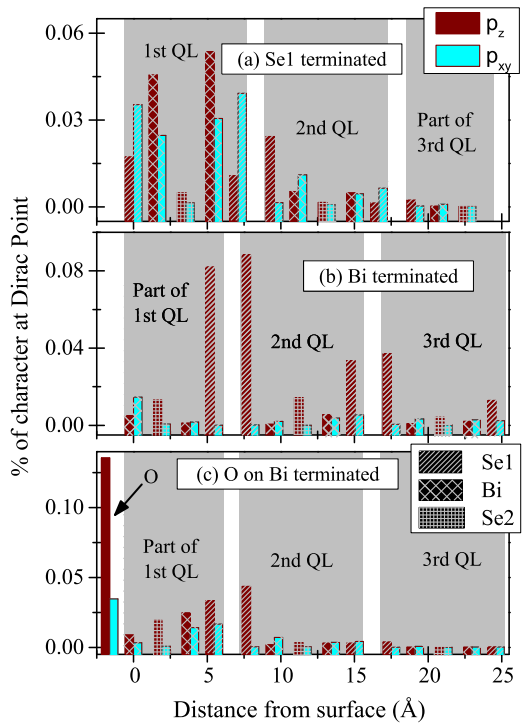}  %trim option's parameter order: left bottom right top
 \vspace{-2ex}
\caption{Contribution to the Dirac point from different atomic
layers. (a) S3, (b) S4 and (c) O on S4 terminations.}
 \vspace{-2ex}
\end{figure}

In order to understand the above scenario better, we investigate the
character of the Dirac states in Fig. 5. The conduction band in this
system consists of Se 4$p$ and Bi 6$p$ states. The Se 4$p$ and Bi
6$p$ contributions to the Dirac point (intensity within $\pm$50~meV
across the Dirac point) obtained for different layers of Se1
terminated surface is shown in Fig. 5(a). The major contribution
arises due to the hybridized states of Bi $p$ and Se1 $p$ states
situated on both sides of the first quintuple layer
\cite{spin_texture}. Interestingly, the contribution from the
interface of the first and second quintuple layers is most
significant. The Dirac states consisting of Se1 ($p_x$,$p_y$) and Bi
$p_z$ states provide the largest contribution indicating their
surface nature that is linked to the bulk via hybridization. In
addition, there is significant contribution from Se1 $p_z$-Bi
($p_x,p_y$) hybridized states. It is evident that the Dirac states
in this system are not purely surface states, they are contributed
by both surface and interface states involving top quintuple layer.

Ironically, the details of the Bi terminated case depicted in Fig.
5(b) exhibit an unusual scenario of Dirac states arising essentially
from the interface states of first and second quintuple layers. This
is different from the predicted surface behavior and demonstrate the
complex nature of real materials. Here, the Se1 $p_z$ states at the
interface are strongly hybridized and form the Dirac states. In both
the above cases, the Se layers at the interface of the successive
quintuple layers appear to be quite strongly coupled. This explains
why cleaving of these materials require strong force, thereby,
employment of top-post removal method.

The presence of a layer of oxygen on Bi-terminated surface leads to
significant change in the characteristics of the Dirac states. The
Dirac states exhibit dominant surface character as shown in Fig.
5(c) along with significant reduction of the interface
contributions. The large electronegativity of oxygen compared to all
other constituent elements leads to an effective pulling of the
electron cloud toward the surface. These results unambiguously
demonstrate a way to engineer topological order on real systems,
which can remain stable for long and can be used for various device
application, realization of novel physics etc.

In summary, the density functional calculations of Bi$_2$Se$_3$
exhibit varied electronic structure on differently terminated
surfaces. Although energy bands for every surface termination
indicate cases of strong topological order, the signature of Dirac
cone appears on Se1, Se2 and Bi terminated surfaces as defined by
S1, S3 and S4 cases. We discover that the Dirac states are
contributed by both sides of the top quintuple later with dominant
contribution arising from the interface states rather than surface
states of the pristine sample. Oxygenation leads to a shift of the
Dirac point to higher binding energies for Se-terminated surface and
makes the topological order weak. However, the Bi terminated surface
is most robust and evolves to a better defined topological order.
Robustness of topological surface states are the key factor for the
technological applications of these materials and realization of
various exotic physics such as Majorana Fermions, magnetic monopole
etc. These results demonstrate ways of engineering Dirac states on
real materials, which lays the all important building block in this
field.


\begin{thebibliography}{99}
%
\bibitem{hasan_rev} M. Z. Hasan and C. L. Kane, Rev. Mod. Phys. {\bf 82},
3045 (2010).
%
\bibitem{3dti} L. Fu, C. L. Kane, and E. J. Mele, Phys. Rev. Letts.
{\bf 98}, 106803 (2007).
%
\bibitem{majorana} L. Fu and C. Kane, Phys. Rev. Letts. {\bf 100}, 096407 (2008).
%
\bibitem{monopole} X.-L. Qi, R. Li, J. Zang, and S.-C. Zhang, Science
{\bf 323}, 1184 (2009).
%
\bibitem{aging_hasan} D. Hsieh {\it et al}., Nature {\bf 460}, 1101 (2009).
%
\bibitem{relaxation_noh} H.-J. Noh {\it et al}., Europhys. Letts.
{\bf 81}, 57006 (2008).
%
\bibitem{dft_fukai} N. Fukui {\it et al}., Phys. Rev. B {\bf 85}, 115426 (2012).
%
\bibitem{BB_bianchi} M. Bianchi {\it et al}., Nat. Commun. {\bf 1}, 128 (2010).
%
\bibitem{dft_wang} X. Wang, G. Bian, T. Miller, and T.-C. Chiang,
Phys. Rev. Letts. {\bf 108}, 096404 (2012).
%
\bibitem{BB_zhang} Z. Zhang and J. T. Yates Jr., Chem. Rev.
{\bf 112}, 5520 (2012).
%
\bibitem{deep} D. Biswas, S. Thakur, K. Ali, G. Balakrishnan,
and K. Maiti, {\it arXiv} 1411.0801v1.
%
\bibitem{aging_kong} D. Kong {\it et al}., ACS Nano {\bf 5}, 4698-4703 (2011).
%
\bibitem{o2_shen} Y. L. Chen {\it et al}., Science {\bf 329}, 659 (2010).
%
\bibitem{contradict_rader} L. V. Yashina {\it et al}., ACS Nano
{\bf 7}, 5181 (2013).
%
\bibitem{stable_atuchin} V. V. Atuchin {\it et al}., Crystal Growth \&
Design {\bf 11}, 5507 (2011).
%
\bibitem{stable_goly} V. A. Golyashov {\it et al}., J. Appl.
Phys. {\bf 112}, 113702 (2012).
%
\bibitem{wien2k} P. Blaha, K. Schwarz, G. K. H. Madsen, D. Kvasnicka,
and J. Luitz, WIEN2k An Augmented Plane Wave + Local Orbitals
Program for Calculating Crystal Properties., K. Schwarz (ed.)
(Techn. Universit\"{a}t Wien, Austria, 2001).
%
\bibitem{lattice} M. I. Zargarova, P. K. Babaeva, D. S. Azhdarova,
Z. D. Melikova, and S. A. Mekhtieva, Inorg. Mater. {\bf 31}, 263
(1995).
%
\bibitem{vesta} K. Momma and F. Izumi, J. Appl. Crystallogr. {\bf 44},
1272 (2011).
%
\bibitem{tdas_stm} Lin Hsin {\it et al.}, Nano Letts. {\bf 13}, 1915
(2013).
%
\bibitem{GGA} J. P. Perdew, K. Burke, and M. Emzerhof, Phys.
Rev. Letts. {\bf 77}, 3865 (1996).
%
\bibitem{yeir2o7_band} K. Maiti, Solid State Commun. {\bf 149}, 1351
(2009); R. S. Singh, V. R. R. Medicherla, K. Maiti, and E.V.
Sampathkumaran, Phys. Rev. B {\bf 77} 201102(R), (2008).
%
\bibitem{bandgap_black} J. Black, E. M. Conwell, L. Seigle, and
C. W. Spencer, J. Phys. Chem. Solids {\bf 2}, 240 - 251 (1957).
%
\bibitem{bandgap_mooser} E. Mooser and W. B. Pearson, Phys. Rev.
{\bf 101}, 492 (1956).
%
\bibitem{band_inversion} H. Zhang {\it et al.}, Nat. Phys. {\bf 5},
438 (2009).
%
\bibitem{symm_slab} K. Park, J. J. Heremans, V. W. Scarola, and D. Minic,
Phys. Rev. Letts. {\bf 105}, 186801 (2010).
%
\bibitem{cab6} K. Maiti, V. R. R. Medicherla, S. Patil, and R. S. Singh,
Phys. Rev. Letts. {\bf 99}, 266401 (2007).
%
\bibitem{cab6band} K. Maiti, Europhys. Letts. {\bf 82}, 67006 (2008).
%
\bibitem{spin_texture} Z.-H. Zhu {\it et al}., Phys. Rev. Letts.
{\bf 110}, 216401 (2013).
%
\end{thebibliography}
\end{document}